\pdfoutput=1

\documentclass[10pt]{article}

\usepackage{listings}    
\usepackage{graphicx}    
\usepackage{subcaption}  
\usepackage{algorithm}   
\usepackage{algorithmic} 
\usepackage{booktabs}    
\usepackage{hyperref}    
\usepackage[htt]{hyphenat} 
\usepackage{microtype} 
\DisableLigatures{}    

\usepackage{caption}
\usepackage{setspace}
\usepackage{multirow}
\usepackage{enumerate}
\usepackage{pdflscape}
\usepackage{pgfplots} 

\usepackage{xcolor}
\definecolor{codegreen}{rgb}{0.25,0.5,0.35}
\definecolor{codegray}{rgb}{0.5,0.5,0.5}
\definecolor{codepurple}{rgb}{0.6,0,0}
\definecolor{backcolour}{rgb}{0.95,0.95,0.92}
\definecolor{colorstring}{rgb}{0.5,0,0.35}
\definecolor{rltred}{rgb}{0.5,0,0}
\definecolor{rltgreen}{rgb}{0,0.5,0}
\definecolor{rltblue}{rgb}{0,0,0.5}
\definecolor{DarkGreen}{rgb}{0.00,0.60,0.00}
\definecolor{ScarletRed}{rgb}{0.80,0.00,0.00}
\definecolor{blizzardblue}{rgb}{0.67, 0.9, 0.93}
\definecolor{green-yellow}{rgb}{0.68, 1.0, 0.18}
\definecolor{dkgreen}{rgb}{0,0.6,0}
\definecolor{gray}{rgb}{0.5,0.5,0.5}
\definecolor{mauve}{rgb}{0.58,0,0.82}
\definecolor{lightgrey}{rgb}{0.90,0.90,0.90}
\definecolor{grey}{gray}{0.75}
\definecolor{light-gray}{gray}{0.80}

\lstdefinestyle{mystyle}{
    escapechar=©, 
	backgroundcolor=\color{backcolour},
    basicstyle=\footnotesize\ttfamily,
   	identifierstyle=\footnotesize\ttfamily,
	commentstyle=\color{codegreen},
	keywordstyle=\color{colorstring}\bfseries,
	numberstyle=\ttfamily\color{codegray},
	stringstyle=\ttfamily\color{DarkGreen},
	breakatwhitespace=false,
	breaklines=true,
	captionpos=b,
	keepspaces=true,
	numbers=left, 
	numbersep=2pt,
	showspaces=false,
	showstringspaces=false,
	showtabs=false,
	tabsize=2
}
\usepackage{xspace}
\newcommand{\evo}{{\sc EvoMaster}\xspace}

\usepackage{boxedminipage}
\newenvironment{result}%
{\smallskip
	\noindent
	\let\emph=\textbf
	\begin{boxedminipage}{\columnwidth}\begin{center}\em}%
		{\end{center}\end{boxedminipage}%
}

\usepackage{ifthen}
\newboolean{showcomments}
\setboolean{showcomments}{true} 

\ifthenelse{\boolean{showcomments}}{
	\newcommand{\nbc}[3]{
		{\colorbox{#3}{\bfseries\sffamily\scriptsize\textcolor{white}{#1}}}
		{\textcolor{#3}{\sf\small$\langle$\textit{#2}$\rangle$}}}
	
}{
	\newcommand{\nbc}[3]{}

}

\usepackage{authblk}                
\usepackage[square,numbers]{natbib} 
\usepackage{amsmath}                

\newcommand{\ARATRL}{{\sc ARAT-RL}\xspace}
\newcommand{\Restler}{{\sc Restler}\xspace}
\newcommand{\Schemathesis}{{\sc Schemathesis}\xspace}
\newcommand{\bbEvo}{{\sc EvoMaster BB}\xspace}
\newcommand{\wbBase}{{\sc EvoMaster WB}\xspace}
\newcommand{\wbMongo}{{\sc Mongo}\xspace}

\usepackage{geometry}
\title{
Search-Based Fuzzing For RESTful APIs That Use MongoDB
}

\author[1]{Hernan Ghianni}

\author[2]{Man Zhang}

\author[1,3,4]{Juan P. Galeotti}

\author[4,5]{Andrea Arcuri}

\affil[1]{Computer Science Dept., School of Exact and Natural Sciences, University of Buenos Aires, Buenos Aires, Argentina}

\affil[2]{School of Computer Science and Engineering, Beihang University, Beijing, China}

\affil[3]{Institute of Computer Science Research, CONICET, Buenos Aires, Argentina}

\affil[4]{School of Economics, Innovation and Technology, Kristiania University of Applied Sciences, Oslo, Norway}

\affil[5]{Department of Computer Science, Oslo Metropolitan University, Oslo, Norway}

\date{}

\begin{document}

\maketitle

\begin{abstract}
In RESTful APIs, interactions with a database are a common and crucial aspect.
When generating white-box tests, it is essential to consider the database's state (i.e., the data contained in the database) to achieve higher code coverage and uncover more hidden faults.
This article presents novel techniques to enhance search-based software test generation for RESTful APIs interacting with NoSQL databases.
Specifically, we target the popular MongoDB database, by dynamically analyzing (via automated code instrumentation) the state of the database during the test generation process.
Additionally, to achieve better results, our novel approach allows inserting NoSQL data directly from test cases.
This is particularly beneficial when generating the correct sequence of events to set the NoSQL database in an appropriate state is challenging or time-consuming.
This method is also advantageous for testing read-only microservices.
Our novel techniques are implemented as an extension of EvoMaster, the only open-source tool for white-box fuzzing RESTful APIs.
Experiments conducted on six RESTful APIs demonstrated significant improvements in code coverage, with increases of up to 18\% compared to existing white-box approaches.
To better highlight the improvements of our novel techniques, comparisons are also carried out with four state-of-the-art black-box fuzzers.

\end{abstract}

{\bf Keywords}: REST API, Fuzzing, Search-based Software Testing, NoSQL Database, Test Case Generation

\section{Introduction}

Nowadays, microservices play a crucial role in our everyday life. 
Many critical domains (e.g., banking, insurance, and healthcare) rely on microservices to provide their day-to-day operations~\cite{jamshidi2018microservices,newman2021building}.
A microservice~\cite{newman2021building} is a specific software design pattern where a large application is divided into smaller, independent \emph{services} that are orchestrated to achieve an overall behaviour.

A popular style to implement such microservices is the Representational State Transfer API (RESTful API or REST API, for short)~\cite{fielding2000architectural}.
Here, microservices achieve interoperability by handling interactions through HTTP(S) calls for seamless communication.
Data, objects or any other meaningful entities are represented as \emph{resources}. 
Each resource is unequivocally identified with a Uniform Resource Identifier (i.e., URIs), and can be manipulated using the semantics of an HTTP call, e.g., \texttt{GET} requests to fetch data, \texttt{POST} to create new data, \texttt{PUT}/\texttt{PATCH} to modify existing data and \texttt{DELETE} to remove data.
Inputs to these requests can be given specifying path elements in the URLs, query parameters, HTTP headers and body payloads. 
There is no a universal format for data transfer, although one of the most common is the JavaScript Object Notation (JSON).

In order to persist data changes to the application, REST APIs usually depend on database systems~\cite{golmohammadi2022testing}.
Structured Query Language (SQL)-based databases offer a relational model to store uniformized data into a table-based structure (rows and columns)~\cite{SQL}.
In contrast, NoSQL databases~\cite{NoSQL} allow storing data into a unstructured, loose, format.
This proved to allow applications both scalability and flexibility. 

Software testing~\cite{orsoTravelogue2014} is the preferable technique for assessing that a given RESTful API complies with its expected behaviour.
However, manually writing test cases to detect mismatches between the intended and the expected output (i.e., defects or \emph{bugs}) is a task both difficult and expensive.
Automatic test case generators~\cite{Candea2019,orsoTravelogue2014} are tools that are specifically tailored for creating test cases on a given program.
The most common automatic test case generators for RESTful APIs are black-box ones~\cite{golmohammadi2023testing}.
More specifically, the inner behaviour of the System Under Test (SUT) is unknown to the automatic test generator during the generation process.
It can only rely on the outputs that are produced on each request by the RESTful API.
In contrast, a white-box RESTful test case generator might profit from runtime information from the SUT, aiming at generating test cases that execute previously non-covered regions of the program. 
At the time of the writing of this article, \evo~\cite{arcuri2018evomaster,arcuri2021evomaster} is the only white-box test case generator for  RESTful APIs, supporting languages that can run on the JVM.
\evo is a search-based testing tool that profits from runtime information.
To guide the test case generation, \evo instruments the SUT to retrieve both the number of covered lines and a computed \emph{branch distance}~\cite{arcuri2019restful} heuristic score for each reached branch, as well as computing other advanced white-box heuristics~\cite{arcuri2021tt,arcuri2024advanced}.

Existing work of handling SQL databases~\cite{arcuri2020sql} has presented
effectiveness to improve coverage and fault detection for fuzzing RESTful APIs.
In a nutshell, it introduces a \emph{secondary} goal for each SQL query that returned no rows.   
The intuition behind this is that, whenever such SQL queries return a non empty result, it might trigger code paths that were previously unexecuted, thus enhancing code coverage and increasing the likelihood of covering undetected failures. 
Therefore, the SUT instrumentation computes a new heuristic score (similar to the branch distance) that approximates how close the database was from satisfying the criteria to return a non empty result when executing the SQL query.  
Additionally, to handle scenarios where data insertion in the database is not feasible (i.e., the RESTful API is \emph{read-only}), or the sequence of interactions with the RESTful API is too complex to craft the appropriate data, \evo is enabled to synthesize and insert data \emph{directly} into the relational database.
 
Inspired by the aforementioned approach, we present in this article a new technique to specifically handle NoSQL databases. 
This approach does not merely repeats the same solution presented in~\cite{arcuri2020sql} in the context of NoSQL database, since these specific databases pose new research challenges, like their lack of a rigid schema throughout the execution of the SUT.
In particular, we focus on MongoDB, which is the most popular NoSQL database, with more than 40 million downloads.\footnote{\url{https://www.mongodb.com/resources/basics/databases/nosql-explained/most-popular-nosql-database}, accessed 11th September 2024.}

In particular, this article provides the following research and engineering novel contributions:
\begin{itemize}
\item We provide novel NoSQL heuristics that can be integrated into search-based test generation tools for system testing.
\item We provide novel techniques to enable the direct generation of NoSQL data, and how they can be integrated with the regular search for the SUT's inputs.
\item To enable the replicability of our experiments, our extensions to \evo and a replication package are released as open-source software, available at \ \url{https://github.com/anonymousmongodb/fuzzing-nosql}.
\end{itemize}
%


\section{Motivating Example}
\label{sec:motivating_example}

\begin{figure}
\begin{lstlisting}[language=java,basicstyle=\footnotesize,numbers=none]
@RestController
@RequestMapping(path = "/api/nosqlrest")
public class NoSQLRest {
 @Autowired private NoSQLRepository r;
 @RequestMapping(method = POST)
 public void post() {
  NoSQLEntity e = new NoSQLEntity();
  e.setX(42);
  e.setY('b');
  e.setZ(1.0);
  r.save(e);
 }
 @RequestMapping(path = "/{x}/{y}/{z}", method=GET,
    produces=APPLICATION_JSON_VALUE)
 public ResponseEntity get(
      @PathVariable("x") int x, 
      @PathVariable("y") char y,
      @PathVariable("z") double z) {
  List<NoSQLEntity> l = r.findByXAndYAndZ(x+42,
     y+1,z/3.0);
  if (l.isEmpty()==false) {
    // true branch
    return ResponseEntity.status(OK).build();
  } else {
    // false branch
    return ResponseEntity.status(NOT_FOUND).build(); 
 }}
}
public interface NoSqlRepository 
    extends MongoRepository<NoSQLEntity, String> {
    
  List<NoSQLEntity> findByXAndYAndZ(int x, char y, double z);
}
\end{lstlisting}

\caption{\label{fig:example} Example of RESTful API in Java SpringBoot, with two endpoints (a
\texttt{GET} and a \texttt{POST}) accessing a database via Spring Data MongoDB.}
\end{figure}

\begin{figure}
\centering
\begin{lstlisting}[language=java,basicstyle=\small,numbers=none]
{ 
	"_id": {"$oid":"e3..ac"}, 
  "x": 42, 
  "y": "b", 
  "z": 1.0 
}
\end{lstlisting}
\caption{\label{fig:json} A JSON representation of the object that is stored in the database.
The \texttt{\_id} field stands for the object identifier,  a 24-character hexadecimal string (12 bytes), which MongoDB uses as the unique identifier for documents within a single collection. 
}
\end{figure}

In this section, we will briefly show by means of an example how a RESTful API can interact with MongoDB, a widely used NoSQL database. 
The example is written in Java, using the Spring Boot framework~\cite{SpringBoot}, with the Spring Data MongoDB~\cite{DataMongoDB} extension for handling persistence.
Figure~\ref{fig:example} presents an excerpt of the class \texttt{NoSQLRest} and interface \texttt{NoSQLRepository}.
We do not show all the classes and needed configuration files to run this API, but just those we consider necessary to understand the presented example.

In this trivial example, we have only two operations \texttt{POST} and \texttt{GET} on two different endpoints: \texttt{/api/nosqlrest/} (for the \texttt{POST} operation) and \texttt{/api/nosqlrest/\{x\}/\{y\}/\{z\}} (for the \texttt{GET} operation).

The \texttt{POST} operation saves a \texttt{NoSQLEntity} into the MongoDB repository.
In order to do so, the Spring MongoDB library translates the \texttt{NoSQLEntity} instance into the Binary JSON (BSON) format, which is the native data format of MongoDB. 
This representation will contain the values of the translated instance for fields \texttt{x}, \texttt{y} and \texttt{z} (namely, \texttt{42}, \texttt{"b"} and \texttt{1.0}).
Figure~\ref{fig:json} presents a human readable version (in JSON format) of the BSON that is stored in the database.
Observe that a field \texttt{\_id} is added with an automatically generated identifier to identify unequivocally the element in the MongoDB database.
By default, the Spring Data MongoDB will create and store all the JSON objects that are saved in the  \texttt{repository} into a \emph{collection} called \emph{``nosqleentities''}.
We could modify the expected collection name that is created and managed by MongoDB using the \texttt{@Document} annotation in the \texttt{NoSQLEntity} class declaration.

The \texttt{GET} operation retrieves a subset of instances of \texttt{NoSQLEntity} that are stored in the underlying \texttt{``nosqleentities''} collection in the MongoDB database.
Given the \texttt{x}, \texttt{y} and \texttt{z} path parameters in the endpoint, the Spring MongoDB library will select all the JSON objects that satisfy that field \texttt{x} is equal to path parameter \texttt{x} plus $42$, \texttt{y} is equal to path parameter \texttt{y} plus one,  and path parameter \texttt{z} is equal to \texttt{z} divided by $3.0$, constructing a \emph{filter} (i.e., a MongoDB query) to match documents where all three fields are equal to  those computed values.
By means of reflection, new \texttt{NoSQLEntity} instances will be built and fields will be modified accordingly to the data fetched from the MongoDB underlying collection.
   
Notice that the code that actually performs the search on the MongoDB collection is not shown.
The \texttt{@Autowired} annotation signals the Spring framework that it must instantiate a \emph{singleton} repository in field \texttt{repository} in order to save and load \texttt{NoSQLEntity} instances from the MongoDB database.
 Given the naming convention that complies the method declared in the \texttt{NoSqlRepository}, Spring discovers that the particular method is intended to fetch data with given values for fields \texttt{x}, \texttt{y} and \texttt{z}. 
 Internally, Spring will produce the following filter $F$ by using the values passed as arguments to that method:
 $$
F = \{ ``x": 82,   ``y": ``b",  ``z": -1.0 \}
 $$
 Filters could be directly written as JSON objects, or they could be constructed programmatically using the \texttt{Filter} API provided by MongoDB. 

In this example, handling the scenario where the returned list is non-empty (i.e., the true branch) presents a recurring challenge for search-based test generators. These techniques rely on heuristic methods to determine whether a test case is \emph{``closer''} to reaching a specific execution path. A common heuristic such as branch distance assigns a fixed non-zero value (representing covering the false branch) until the true branch is covered, at which point a zero value is returned. Essentially, the search remains largely random until it happens upon a test case that triggers the desired execution path.
This recurrent challenge is known as the \emph{flag problem}~\cite{BaS03,HHH02}.
In order to overcome this limitation, the search-based test generator requires satisfying the filter that is being created and applied on the \texttt{nosqlentities} collection in the MongoDB database. 
In other words, it is required to find values for the path parameters such that filter produced when invoking \texttt{findByXAAndYAndZ()} is satisfied.

In this article, we introduce a novel technique to provide gradient when retrieving MongoDB data.
This is done by computing an heuristic score approximating how close are the documents stored in the collection to satisfying a desired filter.
The technique monitors all filters submitted to each collection in the underlying database, recording those that upon execution did not return any element.
After test execution, for each such filter, the heuristic score is computed.
By extending \evo with the presented technique, we are able to generate test cases as the one shown in Figure~\ref{fig:generated_test}.
As Java integers have $2^{32}$ possible values, Java char(acter) values have a domain of $2^{16}$ values and Java floating-point double precision values are $2^{64}$, finding a specific $\langle x,y,z \rangle$ has a uniform probability of 
$
\frac{1}{2^{32} \times 2^{16} \times 2^{64}} = \frac{1}{2^{112}}.
$
Note that in this example, a technique like seeding~\cite{rojas2016seeding}, which could potentially enhance the search space, proves ineffective. This is because the path parameters are altered before being submitted as search values to the MongoDB filter.
 
Additionally, we introduce a technique to handle scenarios where certain operations, such as the \texttt{POST} operation, are unavailable (e.g., when the database is read-only). Unlike relational databases, NoSQL databases like MongoDB allow the storage of objects with varying attributes in the same collection. For instance, if our technique inserts a document into the \texttt{nosqlentities} collection that does not conform to the format in Figure~\ref{fig:json}, all subsequent \texttt{GET} operations will throw an exception signaled in the Spring MongoDB library.
This issue makes dealing with NoSQL databases much harder than dealing with SQL ones, as those have precisely defined schemas.
Due to the lack of a fixed schema, our \evo extension dynamically infers an expected type for the elements in a MongoDB collection at runtime. Figure~\ref{fig:generated_test_with_insertion} shows a test case generated by \evo using our extended Domain-Specific Language (DSL) for inserting documents into MongoDB collections. This test first adds a JSON document with fields \texttt{x=53}, \texttt{y="f"}, and \texttt{z=11.0} to the \texttt{nosqlentities} collection.
Subsequently, the \texttt{findByXAndYAndZ()} method is invoked with the same values (since \texttt{11 + 42 = 53}, \texttt{"f" + 1 = "e"}, and \texttt{33.0 / 3.0 = 11.0}), leading to covering the true branch in the shown example.



\begin{figure}
\begin{lstlisting}[language=java,basicstyle=\footnotesize,numbers=none]
@Test
public void test1 () throws Exception {
  given().accept ( "*/*" )
    .post(baseUrlOfSut + "/api/nosqlrest")
    .then ()
    .statusCode(200);
   
  given().accept ( "*/*" ) 
    .get(baseUrlOfSut + "/api/nosqlrest/0/a/3.0")
    .then()
    .statusCode(200);
}
\end{lstlisting}
\caption{\label{fig:generated_test} A test case generated by EvoMaster that satisfies the true branch in Figure~\ref{fig:example}.}
\end{figure}

\begin{figure}
\begin{lstlisting}[language=java,basicstyle=\footnotesize,numbers=none]
@Test
public void test1 () throws Exception {
  List<MongoInsertionDto> insertions = mongo()
    .insertInto("mymongodb", "nosqlentities")
    .d("{ \"x\": 53, \"y\": \"f\", \"z\": 11.0 }")
    .dtos();
      
  MongoInsertionResultsDto insertionsresult = controller
    .execInsertionsIntoMongoDatabase(insertions);
     
  given().accept ( "*/*" ) 
    .get(baseUrlOfSut + "/api/nosqlrest/11/e/33.0")
    .then()
    .statusCode(200);
}
\end{lstlisting}
\caption{\label{fig:generated_test_with_insertion} A test case generated by \evo that satisfies the true branch in Figure~\ref{fig:example} without using the \texttt{POST} operation.}
\end{figure}

\section{Background}
\label{sec:background}

\subsection{NoSQL - No Structured Query Language Databases}

NoSQL databases~\cite{NoSQL} refer to a collection of non-relational databases designed to handle large volumes of diverse data, with high flexibility and scalability, especially for distributed systems. 
Unlike traditional relational databases (i.e., SQL databases), NoSQL databases do not require a fixed schema and allow data to be stored in a variety of formats, such as key-value pairs, document-based structures, wide-column stores, or graph-based representations.
This flexibility is advantageous for systems where the data structure is dynamic or where schema evolution is frequent.

In NoSQL databases, the traditional concepts of tables, rows, and columns do not apply. 
Instead, data can be stored in more flexible structures. 
MongoDB~\cite{MongoDB} is a document-based database where data is stored as documents in collections. 
A document $D_i$ is encoded in Binary JSON (BSON) and can contain a varied number of key-value pairs, allowing for schema flexibility. 
These documents may contain nested structures and arrays, enabling complex data relationships to be represented within a single document. 
As an example, a document $d$ can contain:
\begin{small}
$$
\begin{aligned}
d = \{ & ``name": ``Foo", ``year": 2025, \\
       & ``location": \{ ``city": ``Oslo", ``country": ``Norway" \}\} \\
\end{aligned}
$$
\end{small}
NoSQL databases support a range of query and data manipulation mechanisms depending on the data model. 
As an example, in MongoDB, the query language allows for complex filtering with commands like \texttt{find()} and aggregation pipelines. These queries can include conditions, projections, and sorting, and aggregation functions for performing computations on collections of documents, such as summing values or averaging fields.

\subsection{Search-Based Software Testing}

Over the years, search-based software engineering has emerged as a highly effective approach to tackle a broad spectrum of software engineering challenges~\cite{harman2012search}, particularly in the realm of software testing~\cite{ABHP09}. Notable tools such as EvoSuite~\cite{fraser2011evosuite}, for unit testing, and Sapienz~\cite{mao2016sapienz}, for Android testing, exemplify the success of these techniques.

Software testing can be casted as an optimization problem, maximizing both code coverage and fault detection in the generated test suites. 
Once a fitness function is established for the testing problem, a search algorithm can explore the solution space, which in this case represents all possible test cases.

Several types of search algorithms exist, with Genetic Algorithms being among the most popular. 
Specialized search algorithms (such as Whole Test Suite~\cite{GoA_TSE12}, MOSA~\cite{dynamosa2017} and MIO~\cite{arcuri2018test}) have been developed for the specific problem of generating test suites (in contrast to solely generating test cases). 

One common heuristic for guiding the search process is the branch distance, which helps find input data that satisfies the constraints in logical predicates of branches~\cite{Kor90}. 
The branch distance for any execution of a predicate is computed using a set of recursively defined rules, as shown in Table~\ref{tab:branchdistance}.
For instance, for the predicate $x \geq 100$ where $x = 42$, the branch distance to the true branch is $100 - 42 + K$, with $K > 0$. 
In practice, calculating the branch distance requires instrumenting each predicate to track the distance during execution. 
For more details on branch distance computation, readers are referred to~\cite{Kor90}.

\begin{table}[!t]
\centering
\caption{
Branch distance for relational operands.
$a$ and $b$ are numeric values, whereas $x$ and $y$ are
boolean predicates/expressions.
}
\begin{tabular}{c l}
\toprule 
Relational operation & Branch distance $\rho$\\
\midrule 
$a>b$ 		& $\rho$: if $a>b$ then 0 else $(b-a)$ \\
$a\geq b$ 	& $\rho$: if $a\geq b$ then 0 else $(b-a)+K$ \\
$a<b$		& $\rho$: if $a<b$ then 0 else $(a-b)+K$ \\
$a\leq b$ 	& $\rho$: if $a\leq b$ then 0 else $(a-b)$ \\
$a=b$ 		& $\rho$: if $a=b$ then 0 else $abs(a-b)$ \\
$x || y$	& $\rho$: $min(\rho(x), \rho(y))$ \\
$x \&\& y$  & $\rho$: $\rho(x) + \rho(y)$ \\     
\bottomrule
\end{tabular}
\label{tab:branchdistance}
\end{table}

\subsection{EvoMaster}

\evo~\cite{arcuri2021evomaster,arcuri2018evomaster} is an open-source tool designed to generate system-level test cases for RESTful APIs~\cite{arcuri2019restful}.
It leverages evolutionary algorithms, such as MIO~\cite{arcuri2018test} and MOSA~\cite{dynamosa2017}.
\evo is composed of two main components: (1) the \emph{core}, which handles the command-line interface, search algorithms, and test case generation; and (2) the \emph{controller} library, which users employ to write configuration classes that instruct \evo on how to start, reset, and stop the SUT.
The controller library also automatically instruments the SUT at startup to collect runtime heuristics, such as the aforementioned branch distance.

\evo generates fully  self-contained test cases in several output formats, such as Java, Kotlin, Javascript or Python.
These tests utilize the \emph{controller} to start, stop, and reset the SUT as needed, via \texttt{@Before} and \texttt{@After} methods. They can be executed directly from an IDE, such as IntelliJ, or integrated into a build system like Maven or Gradle. Each generated test consists of a series of HTTP calls to the SUT, leveraging the widely used RestAssured~\cite{RestAssured} library. \evo is open-source and can be accessed at \url{https://www.evomaster.org}.

\section{Related Work}
\label{sec:related_work}

In the last few years, there has been an increasing interest in the research community on automated testing of REST APIs~\cite{golmohammadi2023testing}.
In the literature, several tools have been proposed, like for example (in alphabetic order):
\ARATRL~\cite{kim2023adaptive},
bBoxrt~\cite{laranjeiro2021black},
\evo~\cite{arcuri2020blackbox},
Morest~\cite{liu2022icse},
RestCT~\cite{wu2022icse},
Restest~\cite{martinLopez2021Restest},
\Restler~\cite{restlerICSE2019},
RestTestGen~\cite{viglianisi2020resttestgen},
and
\Schemathesis~\cite{hatfield2022deriving}.
Different studies have been carried out to compare these tools, like for example as done in two recent empirical studies for existing fuzzers for REST APIs~\cite{zhang2023open,Kim2022Rest}.
Apart from \evo that supports white-box testing, all other tools only support black-box testing.

As far as we know, what presented in this paper is the first approach in the literature aimed at analyzing SUT's interactions with MongoDB databases, and use such information to improve how the fuzzing process is carried out.
The closest work in the literature to what we present in this paper is perhaps what presented in~\cite{arcuri2020sql}, which is aimed at SQL databases.
However, there are major differences and research challenges between dealing with SQL and NoSQL databases.
In particular, the lack of formally defined schemas in MongoDB significantly complicates the runtime analyses.

\section{NoSQL Heuristics}
\label{sec:nosql_heuristics}

When testing a RESTful API, multiple criteria may need to be optimized. 
For instance, we might aim to maximize statement coverage, branch coverage, mutation testing scores, fault detection, or even a combination of these criteria~\cite{rojas2015combining}. 
These metrics can be evaluated by executing test cases in an integrated development environment (IDE) or on a Continuous Integration server. 
To achieve such objectives, a common approach in search-based software testing is to utilize the branch distance~\cite{Kor90}, which provides a gradient for solving constraints within the branch statements of the SUT’s code. 
This requires instrumenting the SUT’s code to collect branch distances. 
However, such heuristics become ineffective when constraints are translated into filters that are submitted to the NoSQL database within a third-party library, such as the Spring Data MongoDB library.
Given the the example shown in Figure~\ref{fig:example}, one might automatically construct such constraints by analyzing both method signatures and names of each found class extending the \texttt{MongoRepository} class.
By inspecting the SUT, we could conclude that, in order to return a non-empty result when calling \texttt{findByXAndYAndZ(int,char,double)} we need to match the values of the arguments to the fields \texttt{x}, \texttt{y} and \texttt{z} in the MongoDB collection. 
Nevertheless, this solution is highly dependant to how (a given version of) Spring Data MongoDB behaves.
Not surprinsingly, other Java/Kotlin Object-Document Mappers offer different mechanisms to query stored data.    
For example, Morphia~\cite{Morphia} requires to build a special \texttt{Query} object with the filters being passed as pairs of string-values. 
Figure~\ref{fig:morphia} shows the equivalent implementation of our motivation example but using the Morphia API to retrieve the \texttt{NoSQLEntity} instead of Spring Data MongoDB.

In this article, we propose an approach that is independent of the Object-Document Mappers (ODMs) or third-party library hiding the querying to the MongoDB database.
Our approach builds on~\cite{arcuri2020sql}, extending and adapting it to tackle the unique challenges associated with handling NoSQL databases.

\begin{enumerate} 
\item We monitor all commands issued to the NoSQL database at the MongoDB driver level, ensuring visibility into every database interaction.

 \item Whenever the SUT performs on a MongoDB collection a \texttt{find()} operation that returns no documents, we calculate a heuristic to estimate how close it was to retrieving non-empty data. 
 This heuristic is based on the evaluation of the boolean predicates in the \emph{filter} passed to the \texttt{find()} operation. 
 
 \item These heuristic values, generated for each call to a \texttt{find()} in a MongoDB collection, are added as secondary objectives for optimization. 
 
 \end{enumerate}

The rest of this section presents a detailed discussion of these three points.

\begin{figure}
\begin{lstlisting}[language=java,basicstyle=\footnotesize,numbers=none]
 Query<NoSQLEntity> q = datastore.createQuery(NoSQLEntity.class)
  .filter("x =", x + 42)
  .filter("y =", y + 1)
  .filter("z =", z / 3.0);
        
 List<NoSQLEntity> l = q.find().toList();
\end{lstlisting}
\caption{\label{fig:morphia} Snippet of the RESTful API  using the Object-Document Mapper Morphia instead of Spring Data MongoDB}
\end{figure}

\subsection{Monitoring NoSQL commands}
\label{sec:monitoring_nosql_commands}

Object-Document Mappers (ODMs) are frameworks that simplify the interaction between applications and NoSQL databases, such as MongoDB. 
Popular ODMs for MongoDB include Spring Data MongoDB~\cite{DataMongoDB}, Morphia~\cite{Morphia}, KMongo~\cite{KMongo} and Micronaut Data MongoDB~\cite{Micronaut}. 
These libraries abstract the complexities of MongoDB's query language and operations, providing a high-level interface for developers. 
They automatically convert user-defined objects into MongoDB documents for storage and retrieval, facilitating interaction between application objects and the database. 
This abstraction significantly simplifies common tasks, such as saving and querying data. 
All these libraries rely on the MongoDB Java drivers~\cite{JDMDB}, which ultimately provides a low-level API to manage database connections and execute queries.
Specifically, when these frameworks retrieve documents from a MongoDB collection, they invoke one of the many overloaded variants of the \texttt{MongoCollection.find()} method. Therefore, by monitoring calls to this family of methods, we can effectively capture all interactions between the application and the MongoDB database.
Following the technique introduced in~\cite{arcuri2024advanced}, we monitor each call to \texttt{MongoCollection.find()} (whether in  the SUT or any third-party library) recording the applied filter.


\subsection{NoSQL Distance}

Given a document $d$ and a filter $F$, we compute the heuristic score $H_d(F)$ by recursively applying the definition presented in Table~\ref{tab:nosql_distance_filter}.
Let us explain this definition by means of an example.
Suppose we have the following filter $F$:
$$
F=\{  ``x": \{ ``\$eq": 17 \}  \}
$$
This filter states that field \texttt{x} equals $17$.
For presentation purposes, let us copy the document $d$ taken from Figure~\ref{fig:json}: 
$$
\begin{aligned}
d = \{ 
    & ``\_id": ``\mbox{$<$auto-generated-Id$>$}", ``x": 42, ``y": ``b", ``z": 1.0 \} \\
\end{aligned}
$$
Based on the definition of the heuristic score $H_d(F)$ for document $d$ and a filter $F$, we need to compute the score $H_d(``x",\{ ``\$eq": 17 \})$. 
Table~\ref{tab:nosql_distance_condition} provides the definition of the heuristic score for a condition (in this case, $C=\{ ``\$eq": 17 \}$) applied to a specific field (here, $f=``x"$).
Following the definition of $H_d(``f",C)$, we deduce that the branch distance $\rho(d[``x"]=17)$ must be calculated.
Therefore, according to the branch distance definitions in Table~\ref{tab:branchdistance}, we find that $H_d(F)=\rho(42=17)=abs(42-17)=25$ .
 
 Now, let us consider a more complex filter $F^{\prime}$:
$$
\begin{aligned}
F' = \{ 
    & ``\$and": \\
        & \quad\quad [ \{ ``x": \{ ``\$eq": 17 \} \}, \\
        & \quad \quad \{ ``y": \{ ``\$eq": ``c" \} \} ]\} 
\end{aligned}
$$
This  filter requires not only  that the field \texttt{x} equals $17$, but also that the field \texttt{y} equals "c".
To compute $H_d(F^\prime)$, we recursively calculate $H_d( \{ ``x": \{ ``\$eq":17 \} \})$ and $H_d( \{ ``y": \{ ``\$eq": ``c" \} \})$.
Instead of simply summing these scores, we apply the normalization function $\nu(x)=\frac{x}{x+1}$ as introduced in~\cite{stvrarcuri13}.
This normalization function ensures that each value is mapped to the range $[0,1]$, preventing any one score from disproportionately dominating the total.
Without this normalization, a much larger score could skew the result, leading to inaccurate conclusions about the relative contributions of each value.
 
As can be seen from the cases considered in Tables~\ref{tab:nosql_distance_filter} and~\ref{tab:nosql_distance_condition}, our prototype already supports most of the filter operators defined by MongoDB~\cite{QueryOperatorMongoDB}.
This includes all comparison, logical, element and array operators.
For string comparisons, we apply the Levenshtein distance~\cite{alshraideh2006search}. 
We have not implemented geospatial nor bitwise operators, although they do not pose a significant challenge besides the limited time available for development.

Each time we detect a call to a \texttt{MongoCollection.find()} method, our instrumentation saves the filter $F$ and the name of the collection $c$ on which the filter $F$ was applied.  
Once test execution has finished, we execute a new find command for each collected filter on the corresponding target collection.
If the target collection is not empty, and the filtering retrieves no documents, we compute a heuristic score $H_d(F)$ on each document $d$ on collection $c$.
Finally, the NoSQL distance that approximates how close are the existing $d\in c$ documents to returning any document on  filter $F$ is computed as:
$$
H_c(F)=min_{d\in c}\{H_d(F)\}
$$

It is worth mentioning that, between the moment in which the filter was observed and the instant in which the NoSQL distance is computed, the database state might have changed (i.e., documents being added, deleted, or updated). 
Although this might lead to an inaccurate gradient, we chose to follow this approach since it provides a practical balance between implementation complexity and performance.

\begin{table}[!t]
\centering
\caption{
NoSQL heuristic score definitions based on filter $F$ for document $d$
}
\begin{small}
\begin{tabular}{l | c}
\toprule 
Filter $F$ & Heuristic score $H_d(F)$ \\
\midrule 
$\{ ``f": cond \}$ 						& $H_d(``f", cond)$ \\
$\{ ``\$and": [cond_1,cond_2,...] \}$ 		&  $\sum_{cond_i}\{\nu(H_d(cond_i))\}$\\ 
$\{ ``\$or": [cond_1,cond_2,...] \}$ 		& $min_{cond_i}\{H_d(cond_i)\}$ \\
$\{ ``\$nor": [cond_1,cond_2,...] \}$  		&  $\sum_{cond_i}\{\nu(H_d(\neg cond_i))\}$ \\
\bottomrule
\end{tabular}
\end{small}
\label{tab:nosql_distance_filter}
\end{table}

\begin{table}
\centering
\caption{
NoSQL heuristic score definitions based on field ``f'' and condition $C$ for document $d$.
}
\begin{small}
\begin{tabular}{l | c}
\toprule 
Condition $C$ & Heuristic score $H_d(``f",C)$ \\
\midrule 
$\{ ``\$eq": v \}$ 					& $\rho(d[``f"]=v)$ \\
$\{ ``\$ne": v \}$  					&  $$\textbf{if} $d[``f"]\neq v$ \textbf{then} $0$ \textbf{else} $K$ \\
$\{ ``\$gt": n \}$ 						& $\rho(d[``f"]>v)$ \\
$\{ ``\$gte": n \}$ 					&  $\rho(d[``f"]\geq v)$\\
$\{ ``\$lt": n \}$ 						& $\rho(d[``f"]<v)$ \\
$\{ ``\$lte": n \}$ 						& $\rho(d[``f"]\leq v)$  \\
$\{ ``\$in": [v1,v2,...] \} $				& $min_{v_i}\{\rho(d[``f"]=v_i)\}$\\
$ \{ ``\$nin": [v1,v2,...] \} $ 				& \textbf{if} $\not\exists_{v_i} d[``f"]=v_i$ \textbf{then} $0$ \textbf{else} $K$\\
$ \{ ``\$mod": [div,rem]\} $ 		&  $\rho (d[``f"]\mbox{ mod }div) =rem)$ \\
$ \{ ``\$exists": true \} $ 				&  $min_{``f_i" \in fields(d)}\{ \rho(``f"=``f_i") \}$ \\
$  \{ ``\$exists": false \} $ 				&  \textbf{if} $``f" \not\in fields(d)$ \textbf{then} $0$ \textbf{else} $K$ \\
$  \{ ``\$size": n \} $ 					&  $\rho( array\_length(d[``f"]) =n )$ \\
$ \{ ``\$type": type \}$					&  $\rho(``type(f)"=``type")$\\
$ \{ ``\$all": [v1,v2,...] \}$ 				&  $\sum_{v_i}\{  min_{w_j \in d[``f"]} \{ \nu(\rho(w_j=v_i))\}   )\}$\\
$ \{``\$not":  cond \}$ 				&  $H_d(``f", \neg cond)$\\
\bottomrule
\end{tabular}
\end{small}
\label{tab:nosql_distance_condition}
\end{table}

\subsection{NoSQL Fitness Function}

Once we have computed a heuristic NoSQL distance for each executed \texttt{find()} command that returns no documents, the next step is to use these distances to enhance the search-based test generation. 
In \evo, for each target goal $g$ (e.g., a branch, a statement or a HTTP status code per endpoint) there is a heuristic value $h(t, g) \in [0,1]$, where $h(t, g)=0$ means the target goal $g$ is not reached when executing test case $t$, and $h(t, g)=1$ means the target goal $g$ is covered when executing test case $t$.
Values in between give heuristic gradient to guide the search toward covering these targets.

Besides the primary heuristic $h(t, g)$, \evo also allows having a secondary heuristic value $s(t, g)$ for each target goal $g$.
When two individuals $t_1$ and $t_2$ have the \emph{same} primary $h(t_1, g)=h(t_2, g)$ value, then the fitness function can break the tie by inspecting the secondary values $s(t_1, g)$ and $s(t_2, g)$  in order to decide which one of the two individuals is fitter for evolution.

In order to compute $s(t,g)$, \evo keeps track of the HTTP call $a_i\in t$ which led to the best value for $h(t, g)$, collecting all secondary SQL distances from $a_i\in t$~\cite{arcuri2020sql}. 
We extended \evo to also include  each computed $H_c(F)$ NoSQL distance as a secondary objective.
Let us call this group of database distances $D_i$.

Once \evo collects the set $D_i$ of secondary database distances, $s(t, g)$ can be computed. 
Let $k^{t}_i$ be the number of database (either SQL or NoSQL) commands executed by the test $t$ 
at HTTP call $i$, where $k^{t}_i \ge |D_i|$, then, $s(t, g)$ is computed as an average but only if the number $k^{t}_i$ of database commands is the same. 
If when comparing a test $t_1$ with a test $t_2$ we have  $k^{t_1}_i \neq k^{t_2}_j$ (where $i$ and $j$ are independent), 
then we rather prefer and select the one with larger $k$, i.e., doing more database operations.

By using $s(t, g)$ as a secondary objective to minimize, the aim is to reward test cases that get closer to having \texttt{find()} operations returning non-empty data sets. 
However, this is just a secondary objective, as we cannot be sure that having data returned by the database will necessarily have a beneficial impact on achieving coverage of $g$.

\section{NoSQL Data Generation}
\label{sec:nosql_data_generation}

To achieve a specific testing goal $g$, such as covering a branch in the SUT, the database may need to be in a specific state. 
For instance, the goal might only be covered if a particular \texttt{find()} command returns the desired data. 
Consequently, reaching this goal may require performing a series of HTTP calls on the SUT to prepare the database appropriately.
As discussed in~\cite{arcuri2020sql} for SQL databases, while this approach can theoretically work, it presents two potential challenges:
\begin{enumerate}
\item Preparing the database might involve executing several HTTP calls on the SUT, which not only complicates the search process but also reduces the clarity of the test. 
\item In some cases, it may even be infeasible to insert the required data into the database using the available SUT endpoints. 
This limitation arises when certain parts of the database are effectively ``read-only'' from the SUT's perspective, particularly when multiple systems access the same database.
\end{enumerate}

To address these challenges, \evo includes the capability to write data directly into relational databases~\cite{arcuri2020sql}. 
For a test case $t$ involving multiple HTTP calls to the SUT, \evo inserts specialized \emph{initialization calls} at the start to configure the relational database state before interacting with the SUT. 
These initialization calls specify not only the target table but also the values to be inserted into each column. 
The mutation of these values follows the same rules applied to mutating values in regular HTTP calls within the test case $t$.

In order to enable data insertion, we extend \evo by adding a new type of specialized initialization call for inserting a \emph{document} in a target MongoDB collection $c$. 
As with SQL databases, \evo will probabilistically add such MongoDB initialization calls whenever a \texttt{find()} command fails to return a result in collection $c$. 
Since NoSQL databases allow the storage of documents with varying attributes within the same collection, the insertion of a document will be generally valid, as no schema is defined.
However, it is often the case that it is expected a specific data format for the documents stored in a given collection.
For example, if a given field in the document has an integer value instead of a string, the mapping from the document to the expected result class will fail when the SUT reads and parses such data.
Such failure to comply to an implicit format might lead to missing significant executions in the SUT.

In order to infer the expected data format, whenever a call to \texttt{MongoCollection.find()} is detected, we also collect the expected result class.
With this information, by means of reflection, our instrumentation derives all the result class' fields and their types.
Therefore, whenever \evo inserts a document in a given collection MongoDB $c$, it probabilistically instantiates a document satisfying the inferred data format.

 
\section{Empirical Study}
\label{sec:empirical_study}

In this paper, we carried out an empirical study to answer the following two research questions:

\begin{itemize}
\item[] {\bf RQ1}: How do our novel techniques improve upon existing white-box fuzzing for RESTful APIs?
\item[] {\bf RQ2}: How do our results compare to the ones of existing black-box fuzzers?
\end{itemize}

\subsection{Artifact Selection}
\label{sec:suts}

To answer our research questions, we needed a set of REST APIs for experimentation that use a MongoDB database.
Our novel techniques are implemented for the JVM (e.g., based on bytecode manipulation).
However, this is just a technical detail, as they could be reimplemented for other programming languages as well (e.g., Go and Python).
Although REST APIs are widely used in industry, especially for microservice architectures, these two constraints (i.e., use of MongoDB and running on the JVM) reduces the available pool of possible APIs that could be found on open-source repositories such as GitHub.

To address this issue, we use all the JVM REST APIs using MongoDB that are present in the latest version of the EMB corpus~\cite{icst2023emb,EMB}.
EMB is a curated set of APIs used for experimentation in software testing research for Web APIs.

\begin{table}[!t]
    \centering
    \caption{\label{tab:sut}
    Descriptive statistics of the employed REST APIs.
    For each SUT,
    \#Endpoints represents the number of declared HTTP endpoints in those APIs,
     \#SourceFiles represents the number of files (i.e., the number of public Java classes) composing such SUT,
    and \#LOCs represents their total number of lines of code (LOC).
    }
   \begin{tabular}{  l r r r }\\ 
\toprule 
SUT & \#Endpoints & \#SourceFiles & \#LOCs  \\
\midrule 
\emph{bibliothek}  & 8 & 33 & 2176\\ 
\emph{genome-nexus}  & 23 & 405 & 30004\\ 
\emph{gestaohospital-rest}  & 20 & 33 & 3506\\ 
\emph{ocvn-rest}  & 258 & 526 & 45521\\ 
\emph{reservations-api}  & 7 & 39 & 1853\\ 
\emph{session-service}  & 8 & 15 & 1471\\ 
\midrule 
\emph{Total} & 324 & 1051 & 84531\\ 
\bottomrule 
\end{tabular} 

\end{table}

Table~\ref{tab:sut} shows some statistics of the selected six APIs used for these experiments.
Full details can be found at~\cite{EMB}, including source code and links to the original GitHub repositories those APIs were collected from.
Note that the reported lines of code (LOCs) is only based on the business logic of these APIs.
They do not count all the third-party libraries used in these APIs (which can be millions of lines of code~\cite{icst2023emb}, e.g., for embedded HTTP servers and development frameworks such as Spring).

Using six APIs is less than what done in large tool comparisons such as
in~\cite{zhang2023open} (20 APIs)
and
in~\cite{Kim2022Rest} (20 APIs).
However, six APIs are more than what used when introducing
\Restler~\cite{restlerICSE2019} (2 APIs) and \evo~\cite{arcuri2017restful} (2 APIs),
and same as Morest~\cite{liu2022icse} (6 APIs).
Still, they are less than what employed for evaluating
\ARATRL~\cite{kim2023adaptive} (10 APIs)
and
\Schemathesis~\cite{hatfield2022deriving} (16 APIs).

While tool comparisons of black-box fuzzers are not influenced by the programming language they are written in, or which type of database (if any) they use, our settings are necessarily more restricted (i.e., JVM and MongoDB).
Still, our case study size is comparable to current research standards.

\subsection{Experimental Settings}

To answer our research questions, we carried out two different set of experiments on the six APIs discussed in Section~\ref{sec:suts}.

In the first set of experiments on \emph{white-box} testing, we compared our novel techniques with a default, base version of \evo.
As far as we know, \evo is the only existing open-source white-box fuzzers for REST APIs, so no comparison with other tools can be done.
In these comparisons, the two techniques are evaluated based on achieved code coverage and detected faults.
To reduce bias, these metrics are computed with \evo's own reporting tooling.
Each experiment was repeated 30 times to take into account the randomness of these algorithms.
Each fuzzing session was left running for one hour.
In total, this took $2 \times 6 \times 30 \times 1h = 15 $ days of computation effort.

In the second set of experiments, we ran four existing black-box fuzzers on these six APIs, using the same experiment settings (i.e., 1-hour fuzzing sessions, with 30 repetitions).
These experiments took $4 \times 6 \times 30 \times 1h =30$ days of computation effort.
The selected fuzzers are
\evo~\cite{arcuri2020blackbox},
 \Schemathesis~\cite{hatfield2022deriving},
 \Restler~\cite{restlerICSE2019}
 and
\ARATRL~\cite{kim2023adaptive}.

Black-box \evo was chosen as it makes easier to compare with white-box testing, as being the same tool this reduces possible side-effects of implementation details impacting the comparisons.
Moreover, it was the tool that gave best results in~\cite{zhang2023open,Kim2022Rest}.
With more than 2 000 stars on GitHub, \Restler and \Schemathesis seem to be the most popular tools among practitioners.
Furthermore, \Schemathesis has shown similar results as \evo in tool comparisons~\cite{zhang2023open}, making it arguably the best black-box fuzzer for REST APIs.
Finally, \ARATRL is a more recent tool (compared to the others), where in an empirical analysis~\cite{kim2023adaptive} it was shown to give better results than \evo and \Restler (as well as better than another fuzzer called Morest~\cite{liu2022icse}).
But, \ARATRL was not compared with \Schemathesis in~\cite{kim2023adaptive}.

Besides these four fuzzers, as discussed in Section~\ref{sec:related_work}, there exist other tools for fuzzing REST APIs,
such as for example
bBoxrt~\cite{laranjeiro2021black},
Morest~\cite{liu2022icse},
Restest~\cite{martinLopez2021Restest},
RestCT~\cite{wu2022icse},
and
RestTestGen~\cite{viglianisi2020resttestgen}.
However, arguably, based on existing empirical studies in the literature (e.g.,~\cite{zhang2023open,Kim2022Rest,kim2023adaptive}), the selected four fuzzers are a good representation of the current state-of-the-art in \emph{black-box} fuzzing of REST APIs.

For the experiments, we downloaded the latest released versions of these tools, as of the 6th of August 2024.
This means we used \evo version 3.0.0,
\Schemathesis version 3.33.3
and \Restler version 9.2.4.
For \ARATRL, we did not use its latest published version 0.1 (which is old, from summer 2023),
but rather from its latest commit (code 04e22bb).
This is because the critical option of specifying for how long to run the fuzzing sessions (e.g., for one hour) was not present in the released version 0.1.

These tools were compared based on achieved code coverage.
This was done by instrumenting the APIs with JaCoCo~\cite{JaCoCo}, and collecting code coverage results at the end of 1-hour fuzzing sessions.
However, this approach has two potential issues.

The first issue is that we are not comparing fault detection.
The reason is that each tool report detected faults in their own formats and ways which are not directly comparable.
Looking at returned 500 HTTP status codes via a proxy (e.g., as done in~\cite{corradini2021restats}), or at stack-traces in the server-side error logs (as done in~\cite{kim2023adaptive}) have their own set of limitations.
Furthermore, they only show one type of faults that these tools can detect (e.g., crashes).
But, there are many other kinds of faults that can be detected, e.g., based on schema validation, robustness and security testing~\cite{golmohammadi2023testing}.
Only focusing on one type of faults would be quite restrictive, and possibly biased.
Looking at the total number of self-reported detected faults would be problematic as well, as tools that are equipped with more oracles would be unfairly advantaged.
As such, as long as there is no standarized way among tools on how to report detected faults, focusing only on achieved code coverage is a reasonable compromise for comparisons among different tools.

The second issue is that this approach of measuring code coverage during fuzzing (which is the common approach done in all comparisons in the literature, as far as we know) ignores the generated tests.
A black-box fuzzer could evaluate hundreds of thousands of HTTP calls within one hour.
Generating test suites with hundreds of thousands of HTTP calls would impractical.
Therefore, tools use minimization strategies when outputting test suites, based for example on black-box coverage metrics~\cite{martin2019test}.
However, whether such minimized test suites would still achieve the same level of code coverage is not guaranteed.
Comparing tools based on generated tests is a major technical effort, as tests can be generated in different languages (e.g., Python and JUnit), which would then need to be possibly compiled (with all needed third-party libraries) before being executed.
Furthermore, unless the size of the generated test suites is taken into account in the comparison metrics, tools that output lots and lots of test cases (which would not be useful for practitioners, as not manageable) could have an unfair advantage.

Our novel techniques, aimed at MongoDB support, are for white-box testing.
To compare them with existing black-box fuzzers, we need to use the same metrics, collected with the same tools.
Unfortunately, it would not be recommended to run the APIs with JaCoCo while white-box fuzzing them, as the bytecode manipulations of \evo's instrumentation could conflict with the one of JaCoCo, possibly leading to messed up results.
For the sake of these experiments, we run the tests generated instrumented with JaCoCo \emph{after} the one hour fuzzing sessions.
Note that \evo white-box instrumentation is only needed during the fuzzing (e.g., to collect and compute search-based heuristics).
It is not needed for when the generated tests are run, so there is no conflict of any kind with JaCoCo.
As \evo's minimization algorithms take into account the achieved code coverage (as that is computed for each test execution), there is no risk of losing code coverage in the generated test suites (unless there is a fault in \evo itself).
 However, how \evo measures code coverage and how JaCoCo does that is not necessarily the same.

These two sets of experiments took together $15 + 30 = 45$ days of computation effort.
All experiments were run on a HP Z6 G4 Workstation with Intel(R) Xeon(R) Gold 6240R CPU @2.40GHz 2.39GHz, 192 GB RAM, with 64-bit Windows 11.

\subsection{Experimental Results}

\begin{table*}
	\centering
	\caption{
	Performance comparisons between the \emph{Base} and \emph{Mongo} configurations, in terms of average (i.e., arithmetic mean) line coverage and average number of detected faults.
	Results of statistical tests are reported, including p-values and $\hat{A}_{12}$ effect sizes.
	For p-values lower than the threshold $\alpha=0.05$, the effect sizes $\hat{A}_{12}$  are shown in bold.
	We also report the average number of HTTP calls done during the search, and their scaled difference compared to the \emph{Base} configuration.
	}
	\label{tab:wb}
	\resizebox{1.\linewidth}{!}{
		\begin{tabular}{ l rrrr rrrr rrr}\\ 
\toprule 
SUT & \multicolumn{4}{c}{Line Coverage \%} & \multicolumn{4}{c}{\# Detected Faults} & \multicolumn{3}{c}{\# HTTP Calls} \\ 
    & Base & Mongo  & $\hat{A}_{12}$ & p-value  & Base & Mongo & $\hat{A}_{12}$ & p-value & Base & Mongo & Ratio \\
\midrule 
\emph{bibliothek} & 27.0 & 31.2 & {\bf 1.00} & $< 0.001$ & 0.0 & 2.5 & {\bf 1.00} & $< 0.001$ & 278002 & 133799 & 48.13\\
\emph{genome-nexus} & 37.3 & 38.9 & {\bf 0.86} & $< 0.001$ & 0.0 & 0.0 & 0.50 & 1.000 & 16576 & 12180 & 73.48\\
\emph{gestaohospital-rest} & 40.1 & 53.9 & {\bf 1.00} & $< 0.001$ & 1.0 & 1.7 & {\bf 0.83} & $< 0.001$ & 61829 & 71091 & 114.98\\
\emph{ocvn-rest} & 28.9 & 29.1 & 0.62 & 0.114 & 297.3 & 299.0 & 0.53 & 0.743 & 100824 & 100665 & 99.84\\
\emph{reservations-api} & 49.1 & 55.1 & {\bf 1.00} & $< 0.001$ & 4.0 & 4.9 & {\bf 0.79} & $< 0.001$ & 45581 & 41305 & 90.62\\
\emph{session-service} & 57.3 & 75.7 & {\bf 1.00} & $< 0.001$ & 8.5 & 8.7 & 0.55 & 0.441 & 285487 & 188874 & 66.16\\
\midrule 
Average  & 40.0 & 47.3 & 0.91 &  & 51.8 & 52.8 & 0.70 &  & 131383 & 91319 & 82.20\\
Median  & 38.7 & 46.4 & 1.00 &  & 2.5 & 3.7 & 0.67 &  & 81327 & 85878 & 82.05\\
\bottomrule 
\end{tabular} 

	}
\end{table*}

Table~\ref{tab:wb} shows the results of our first set of experiments,
where the default version of \evo (named \emph{Base} here) is compared with our extension for MongoDB support (named \emph{Mongo}).
The two configurations are compared following standard guidelines in software engineering research~\cite{Hitchhiker14}.
In particular, pair comparisons per SUT are analyzed with Wilcoxon-Mann-Whitney U-tests, with $\hat{A}_{12}$ standarized effect-sizes.

Regarding achieved code coverage,
in all cases but one (i.e., \texttt{ocvn}) our novel techniques provided better results that are statistically significant, with strong effect-sizes (average effect-size $\hat{A}_{12}=0.91$).
On four out of six APIs, the effect-size was the maximum $\hat{A}_{12}=1.00$.
This means that, on these four APIs, every single run out of 30 for \emph{Mongo} gave better results that any run for \emph{Base}.

On the one hand, on \texttt{session-service} the achieved code average increased on average by 18\%.
On the other hand,
the case of \texttt{ocvn} is rather particular.
It seems our novel techniques provide some improvement, but 30 runs are not enough to provide enough confidence to claim it with statistical significance.
Looking at the source code of \texttt{ocvn},
for most of this API's endpoints, data seems that is mainly written into the MongoDB database, and not read.
This could explain why our novel techniques do not provide much benefits for fuzzing an API like this.


Regarding fault detection, based on the results shown in Table~\ref{tab:wb}, our novel techniques can reliably find new faults in three of these APIs (average $\hat{A}_{12}=0.70$).
However, whether faults are found depends entirely on whether such faults exist in the first place.
These six APIs are taken as they are, with no manually injected faults.
Our novel techniques enable to better explore the execution paths of these APIs (see previously discussed improvements on achieved code coverage), but many faults in REST APIs are typically located in the first validation layer~\cite{marculescu2022faults}.
Therefore, proportionally the number of new faults that can be detected would be expected to be lower.

Our novel techniques have a computational cost.
Such cost makes each test case evaluation longer.
Given the same amount of time, this means that less test cases can be evaluated.
If the novel techniques provide no benefit to the search process, then they would have a detrimental effect on the final results.
 In Table~\ref{tab:wb}  we can see that on average there is a decrease of around 18\% in the number of evaluated HTTP calls (note that each test case is composed of one or more HTTP calls towards the tested API).
 However, this is not as simple as stating that this is due to computational overhead, alone.
 The reason is that not all test case executions take the same amount of time.
 A test case that returns immediately due to input validation (e.g., a 400 user-error code) will be faster than a successful call that executes large parts of the API's source code, including calls to databases and external services.

 An example of this is \texttt{session-services}, where only 2/3 of test cases could be evaluated within the same amount of time (i.e., 285k vs.~188k HTTP calls).
 However, much better coverage results were obtained for this API.

\begin{result}
{\bf RQ1}: Our novel techniques provide strong improvements in terms of code coverage compared to the state-of-the-art in white-box fuzzing of REST APIs.
The average effect-size is $\hat{A}_{12}=0.91$, with improvements of up to 18\%.
In terms of fault detection, more faults are found.
\end{result}

\begin{table*}
	\centering
	\caption{
	Performance comparisons between the six analysed tools.
    For each tool, we report the achieved average (out of 30 runs) line coverage (measured with JaCoCo),
     as well as the minimum and maximum values (out of 30 runs) achieved (shown in squared `[]' parentheses).
     For each tool, we report in `()' parenthesis its rank (from `1' best, to `6' worst).
	}
	\label{tab:bb}
	\resizebox{1.\linewidth}{!}{
		\begin{tabular}{ l  r r r r r r    }\\ 
\toprule 
SUT  &  \ARATRL &  \evo BB &  \Restler &  \Schemathesis &  \wbBase &  \wbMongo \\
\midrule 
\emph{bibliothek} & 35.2 [35.2,35.2] (5.5) & 39.7 [39.7,39.7] (3) & 35.2 [35.2,35.2] (5.5) & 39.7 [39.7,39.7] (3) & 39.7 [39.7,39.7] (3) & {\bf 46.9 [42.3,48.3] (1)} \\ 
\emph{genome-nexus} & 12.8 [12.8,12.8] (6) & 31.4 [28.6,33.9] (3) & 24.4 [24.4,24.6] (5) & 27.3 [26.3,28.5] (4) & 35.6 [33.5,36.7] (2) & {\bf 37.4 [31.5,39.4] (1)} \\ 
\emph{gestaohospital-rest} & {\bf 58.0 [45.3,65.6] (1)} & 56.6 [46.6,61.6] (2) & 21.5 [21.5,21.5] (6) & 51.4 [48.0,56.9] (4) & 38.9 [38.7,38.9] (5) & 53.7 [50.3,56.6] (3) \\ 
\emph{ocvn-rest} & 10.1 [10.1,10.1] (5.5) & 27.7 [27.7,27.9] (3) & 10.1 [10.1,10.1] (5.5) & 27.4 [27.3,27.5] (4) & 27.9 [27.8,28.0] (2) & {\bf 28.2 [27.8,34.3] (1)} \\ 
\emph{reservations-api} & 26.2 [26.2,26.2] (6) & 27.6 [27.6,27.6] (4.5) & 27.6 [27.6,27.6] (4.5) & 34.3 [30.8,56.3] (3) & 62.4 [62.4,62.4] (2) & {\bf 70.0 [68.8,70.3] (1)} \\ 
\emph{session-service} & {\bf 88.3 [88.1,89.3] (1)} & 54.1 [54.1,54.1] (4.5) & 50.9 [50.9,50.9] (6) & 54.1 [54.1,54.1] (4.5) & 58.5 [57.9,59.1] (3) & 80.0 [76.7,84.9] (2) \\ 
\midrule 
Average  & 38.4 (4.2) & 39.5 (3.3) & 28.3 (5.4) & 39.0 (3.8) & 43.8 (2.8) & 52.7 (1.5) \\ 
\hline 
Friedman Test & \multicolumn{6}{r}{\textbf{$\chi^2$ = 15.421, $p$-value = 0.009} } \\ 
\bottomrule 
\end{tabular} 

	}
\end{table*}

Table~\ref{tab:bb} shows the results of our second set of experiments.
In this table, six tools are compared:
four black-box fuzzers (i.e., \ARATRL, \bbEvo, \Restler and \Schemathesis),
white-box \evo (i.e., \wbBase),
and \evo extended with our novel techniques (i.e., \wbMongo).
Comparisons are based on achieved code coverage (measured with JaCoCo).
For each API, each tool is ranked based on their performance, where Rank 1 is best, and Rank 6 is the worst.
To verify the difference in performance among the tools, a Friedman Test is carried out.

Table~\ref{tab:bb} shows some interesting results.
First of all, considering the \textit{Average} scores, our novel techniques (i.e., \wbMongo) gives the best overall results (i.e., the best average Rank 1.5),
by a \emph{large} margin in terms of line coverage (i.e., $52.7 - 43.8 = +8.9$\% over the second best tool).
The fact that \wbBase is the second best tool is not surprising.
White-box techniques that can exploit internal knowledge of the tested SUT (e.g., runtime line coverage and processed database commands) have a clear advantage over black-box techniques.
However, surprisingly, on 2 APIs it is \ARATRL that gives the best results (i.e., on \texttt{gestaohospital} and on \texttt{session-service}).
Let us discuss these two cases in more details.

Based on the results in Table~\ref{tab:bb},
with an average code coverage of 58\%, \ARATRL gives better results than the second best tool, which is \bbEvo with average coverage 56.6\%.
Interestingly, on this API \wbBase gives significantly worse results than \bbEvo (i.e., 38.9\%).
To explain this rather odd behavior, we can think of at least two possible  conjectures:
(1) the fitness function for the evolutionary search of \evo gives little to no guidance on this API.
As a black-box approach can evaluated many more test cases in the same amount of time (as it does not have to collect coverage results and compute white-box heuristics at each test case execution), it can better explore the search space.
(2) There is a fault in \evo, where the outputted generated JUnit test cases do not contain all the coverage improving test cases evolved during the fuzzing session.
Recall that, for \wbBase and \wbMongo, coverage is based on the execution of the generated JUnit test suites, and not on all the calls done during the 1-hour fuzzing sessions.
If we were to apply the same approach for \ARATRL, then its results would be 0\% coverage, as that tool prototype does not generate any executable test case (only log summaries).

\texttt{session-service} is another interesting case.
If we exclude the results of \ARATRL, then \wbBase gives better results than any black-box fuzzer,
and \wbMongo provides strong improvements upon it (i.e., from 58.5\% to 80\%).
Still, this is lower than the 88.3\% achieved by \ARATRL.
Fuzzing a REST API poses many variegated challenges~\cite{zhang2023open}.
Handling MongoDB interactions is only one of many aspects that a fuzzer needs to deal with.
In the particular case of \texttt{session-service}, there is something special there that \ARATRL is very efficient at dealing with.
However, without in-depth analyses, at this point in time we cannot really conjecture what that might be.

The goal of this empirical study is to compare our novel techniques with the state-of-the-art, to demonstrate that they provide a useful research contribution.
However, by comparing existing black-box fuzzers, we can look at the existing published literature to see how our replicated comparisons fare on our selection of APIs used for experimentation.
For example,
in the case of the large set of experiments in~\cite{zhang2023open}, our results are consistent:
\bbEvo and \Schemathesis provide the best results, better than \Restler (\ARATRL did not exist yet at that time).
Regarding the work in~\cite{kim2023adaptive} that introduced \ARATRL, our results are quite different.
In our study, \ARATRL still gave better results than \Restler, but not better than \bbEvo (note that \Schemathesis was not compared in~\cite{kim2023adaptive}).
However, on two out of six APIs in our study, \ARATRL gave the best results, even better than white-box testing.

This is an important reminder about what kind of conclusions can be drawn from this type of tool comparisons.

First, the outcome of the comparisons largely depends on the selected APIs used in the experiments.
By cherry-picking SUTs, one can show whatever results they want.
Trying to be as fair and unbiased as possible in the SUT selection is of paramount importance for sound scientific studies.
To provide stronger results, at posteriori we could have excluded from our study  \texttt{gestaohospital} and \texttt{session-service}.
But, besides being unethical and scientifically fraudulent, we could not have claimed then that we used all SUTs in the EMB corpus that rely on a MongoDB database.
Having an established corpus of APIs like EMB is essential for scientific research, to avoid this kind of issues.
However, as for any benchmark, it is important that novel techniques are not tailored for a specific corpus (i.e., no overfitting).
Corpora that are extended each year with new APIs for experimentation is necessary.

Second, ideally, from a scientific standpoint one would like to compare ``techniques'' and not ``tools''.
A tool could give better results just because it is more mature and robust.
Building an effective fuzzer that can handle all the different aspects of OpenAPI schemas and properties of REST APIs is no trivial task.
It requires time and effort.
In this regard, older but actively maintained tools such as \Schemathesis and \evo have a clear advantage.
Still, novel techniques can be designed that can be particularly effective on some APIs, like it happened in our experiments with \ARATRL.
To better ``sell'' our novel results, at posteriori we could have excluded the comparisons with \ARATRL, and rather use some other tools that seem to give worse results (e.g., based on previously published tool comparisons in the literature such as in~\cite{zhang2023open}).
Apart from being unethical, it is our opinion that it would \emph{reduce} the scientific contribution of this paper.
Showing that a novel technique works effectively on some SUTs is important, as well as it is important to show where it does not work so good.
We integrated our novel techniques in the evolutionary search-based engine of \evo.
But, based on the results obtained by \ARATRL, it is clear that there is an important research direction in studying how to integrate these novel techniques in reinforcement learning approaches, like done for example in \ARATRL.
This will be an important future work.

\begin{result}
{\bf RQ2}: on average, with six real-world APIs, our novel techniques provide the overall best results compared to all the analyzed fuzzers in our empirical study.
However, there are important research directions for further improvements, for example on how to integrate our techniques with reinforcement learning approaches.
\end{result}

\section{Threats to Validity}
\label{sec:threats_to_validity}

A threat to internal validity is that the compared tools are using randomized algorithms.
To deal with this, all experiments were repeated 30 times, and analyzed with the appropriate statistical tests recommended in the literature~\cite{Hitchhiker14}, such as for example Wilcoxon-Mann-Whitney U-Test, Vargha-Delaney $\hat{A}_{12}$ standarized effect-size, and Friedman Test.

For external validity, six real-world APIs were used in our empirical study, chosen from the EMB Corpus~\cite{icst2023emb,EMB}.
How good our novel techniques would fare on different APIs is not something that can be known for sure without further experiments.
Using a larger set of APIs for experimentation is not simple, as we are constrained on the programming language (i.e., must run on the JVM, as we do white-box testing based on \evo's bytecode instrumentation) and type of API (it needs to use MongoDB).
Although Java and MongoDB are widely used in industry, finding non-trivial instances of this type of web service on open-source repositories such as GitHub is not so simple.
Collaborations with industrial partners to access to more APIs for experimentation will be an important research direction to follow.

\section{Conclusion}
\label{sec:conclusion}

In this paper, we have proposed a series of novel techniques to handle interactions with MongoDB when doing white-box fuzzing of REST APIs.
Based on automated bytecode instrumentation, we can automatically intercept and analyze at runtime all calls made by the APIs towards their MongoDB databases.
This is done to be able to stir the fuzzing process to generate more effective test inputs, as well as being able to directly insert all the needed data directly into MongoDB from the test cases.

Our novel approach is evaluated on six real-world APIs, providing better results than existing fuzzers, including \evo, \Schemathesis, \ARATRL and \Restler.
Compared to the state-of-the-art, large improvements in code coverage are achieved (e.g., up to $+18$\%), as well as being able to detect more faults on average.

Our empirical study also shows some potential directions for further research to obtain better results.
For example, on some APIs, reinforcement learning techniques as done in \ARATRL give better results.
How to best integrate our novel techniques with reinforcement learning will be an important research question to address.




%


\section*{Acknowledgments}
This work is funded by the European Research Council (ERC) under the European Union’s Horizon 2020 research and innovation programme (EAST project, grant agreement No. 864972), and partially funded by UBACYT-2020 20020190100233BA, PICT-2019-01793.
Man Zhang is supported by State Key Laboratory of Complex \& Critical Software Environment (SKLCCSE, grant No. CCSE-2024ZX-01).

\section*{Data Availability}

\evo and its new extension presented in this paper are open-source on GitHub,\footnote{https://github.com/WebFuzzing/EvoMaster}
with each release automatically uploaded to Zenodo for long-term storage (e.g.,~\cite{zenodo340evomaster}).
To enable the replicability of our experiments, our tool and a replication package are released as open-source software, available at \ \url{https://github.com/anonymousmongodb/fuzzing-nosql}.


\bibliographystyle{ACM-Reference-Format} 

%

\bibliography{papers_mongo.bib}


\end{document}